\documentclass[prd,preprint,superscriptaddress,showpacs,byrevtex]{revtex4}
\usepackage{bm}
\def\fsl#1{\setbox0=\hbox{$#1$}           % set a box for #1
   \dimen0=\wd0                                 % and get its size
   \setbox1=\hbox{/} \dimen1=\wd1               % get size of /
   \ifdim\dimen0>\dimen1                        % #1 is bigger
      \rlap{\hbox to \dimen0{\hfil/\hfil}}      % so center / in box
      #1                                        % and print #1
   \else                                        % / is bigger
      \rlap{\hbox to \dimen1{\hfil$#1$\hfil}}   % so center #1
      /                                         % and print /
   \fi}                                         %
\newcommand{\be}{\begin{equation}}
\newcommand{\ee}{\end{equation}}
\newcommand{\bea}{\begin{eqnarray}}
\newcommand{\eea}{\end{eqnarray}}
\newcommand{\beq}{\begin{equation}}
\newcommand{\eeq}{\end{equation}}
\newcommand{\beqs}{\begin{eqnarray}}
\newcommand{\eeqs}{\end{eqnarray}}

\begin{document}
\title{ Is Renormalization in QCD Necessary at High Energy Colliders ?}
\author{Gouranga C Nayak }\thanks{G. C. Nayak was affiliated with C. N. Yang Institute for Theoretical Physics in 2004-2007.}
\affiliation{ C. N. Yang Institute for Theoretical Physics, Stony Brook University, Stony Brook NY, 11794-3840 USA}
%
%\date{\today}
\begin{abstract}
In this paper by using the path integral formulation of the background field method of QCD in the presence of SU(3) pure gauge background field we simultaneously prove the renormalization of ultra violet (UV) divergences and the factorization of infrared (IR) and collinear divergences in QCD at all orders in coupling constant. We prove that although perturbative QCD is renormalizable but due to confinement in QCD it is not necessary to renormalize QCD to study hadrons production at high energy colliders. This is consistent with the fact that the partons are not directly experimentally observed but the hadrons are directly experimentally observed.
\end{abstract}
\maketitle
\pagestyle{plain}
\pagenumbering{arabic}
\section{Introduction}

Quark and gluon are the fundamental particles of the nature which exist inside the hadron (inside the proton and neutron etc.). Quantum chromodynamics (QCD) is the fundamental theory of the nature which describes the interaction of quarks and gluons. After the discovery that the Yang-Mills theory is renormalizable \cite{hf} and after the discovery of asymptotic freedom in QCD \cite{gs} the perturbative QCD (pQCD) calculation at the partonic level has been extensively studied at high energy colliders.

In the loop calculation in perturbative quantum field theory one encounters ultra violet (UV) divergence when
the momentum integration limit goes to infinity. This UV divergence prevents us to make any physical prediction
as such physical prediction will be divergent. The renormalization program is introduced to handle this UV divergence
to make physical prediction. Under the renormalization program one assumes that the quantities present in the
original lagrangian (unrenormalized quantities) are divergent and are not physical. One defines renormalized quantities which are finite in
terms of which the physical quantities are expressed. The unrenormalized quantities are related to the corresponding
renormalized quantities via divergent constants which cancel the divergences present in the unrenormalized quantities
making the renormalized quantities finite.

The physical interpretation of renormalization in perturbative quantum electrodynamics (QED) is that the experiments
do not measure the unrenormalized quantities because an electron is surrounded by virtual electron-positron
pairs created from the vacuum (loop-diagrams). Hence in the renormalization program one redefines the definition
of the physical quantities to be renormalized quantities whenever quantum corrections (loop diagrams) are included.
Under the renormalization program in perturbative QED what we directly experimentally measure are the renormalized
quantities.

One extends such renormalization program from perturbative QED to perturbative QCD by extending U(1)
gauge theory to SU(3) gauge theory \cite{hf,gs}. It is shown that the perturbative QCD is renormalizable
\cite{hf}. However, the main difference between QED and QCD is that
while we do directly experimentally observe the electron in QED but we do not directly experimentally observe the quark in QCD because of confinement in QCD, a phenomena which is absent in QED. From the experimental point of
view the main difference between QED and QCD is that in QED the experiments directly observe electron which is
the fundamental particle in QED but in QCD the experiments directly observe hadron which is not the fundamental
particle in QCD (the fundamental particles in QCD are quark and gluon).

Hence in QED it is clear that since the electron is directly experimentally observed, the renormalized quantities
in QED are experimentally measured. However, in QCD this is less clear because the quark is not directly
experimentally observed due to confinement in QCD. Since experiments observe hadron, it is less clear if the
renormalized quantities in QCD are necessary for this purpose.

Consider for example the hadrons production at high energy colliders. Let us consider the general partonic level scattering process
\bea
p_1+p_2 \rightarrow p'_1+p'_2+...+p'_n
\label{pl}
\eea
where $p_1^\mu, p_2^\mu$ are the four-momenta of the incoming partons and $p'^\mu_1, p'^\mu_2,...,p'^\mu_n$ are the four-momenta of the outgoing partons. The initial (final) partonic state is given by
\bea
|i> = |p_1,p_2>,~~~~~~~~~~~~~~~|f>=|p'_1,p'_2,...,p'_n>
\label{if}
\eea
and the renormalized partonic level cross section ${\hat \sigma}$ is proportional to the renormalized S-matrix (T-matrix) element square at all orders in coupling constant
\bea
{\hat \sigma} \propto |{\cal M}|^2,~~~~~~~~~~|{\cal M}|^2 \propto |<f|i>|^2.
\label{pc}
\eea

Using the factorization theorem in QCD \cite{s1,s2,s3,n1,n2,n3} the hadrons production cross section at high energy colliders is given by
\bea
{\sigma} = f_1 \otimes f_2 \otimes {\hat \sigma} \otimes D_1 \otimes ....\otimes D_n
\label{sg}
\eea
where $f$ is the renormalized parton distribution function (PDF) inside hadron, $D$ is the renormalized parton to hadron fragmentation function (FF) and the symbol $\otimes$ represents the necessary folding with the PDF (FF).

In this paper we will prove that
\bea
&&{\sigma}=f_1^{Renormalized} \otimes f_2^{Renormalized} \otimes {\hat \sigma}^{Renormalized} \otimes D_1^{Renormalized} \otimes ....\otimes D_n^{Renormalized} \nonumber \\
&&= f_1^{UnRenormalized} \otimes f_2^{UnRenormalized} \otimes {\hat \sigma}^{UnRenormalized} \otimes D_1^{UnRenormalized} \otimes ....\otimes D_n^{UnRenormalized}\nonumber \\
\label{sgu}
\eea
which proves that the UV divergences in the unrenormalized S-matrix element square exactly cancel with the UV divergences in the unrenormalized parton distribution functions and with the UV divergences in the unrenormalized parton fragmentation functions at all orders in coupling constant. This proves that although the perturbative QCD is renormalizable but due to confinement in QCD it is not necessary to renormalize QCD to study hadrons production at high energy colliders.

Eq. (\ref{sgu}) proves that due to confinement in QCD the QCD is a better theory than QED as far as the ultra violet (UV), infrared (IR) and collinear divergences are concerned.

We will provide a proof of eq. (\ref{sgu}) in this paper.

Note that we have derived eq. (\ref{sgu}) by using the symmetry consideration at the Lagrangian level in the path integral formulation of the QCD in this paper without using any Feynman diagrams. Hence our calculation is enormously simplified.

The paper is organized as follows. In section II we show that the infrared/soft divergence arises due to the eikonal part of the Feynman diagram in quantum field theory. In section III we prove that the pure gauge field is produced by the light-like eikonal line in quantum field theory. In section IV we show that the collinear divergence in quantum field theory is described by using the pure gauge field. In section V we derive the relation between the n-point connected green's function in QCD and the n-point connected green's function in the background field method of QCD in the presence of SU(3) pure gauge background field. In section VI we describe the renormalization of the Wilson line. In section VII we discuss the S-matrix element in QCD at all orders in coupling constant and the parton distribution function and the parton to hadron fragmentation function at high energy colliders. In section VIII we present the proof of factorization in QCD at all orders in coupling constant and the cancelation of infrared/soft and collinear divergences in hadrons production at high energy colliders. In section IX we present the simultaneous proof of renormalization and factorization in QCD at all orders in coupling constant at high energy coliders and prove eq. (\ref{sgu}). In section X we briefly review how hadron production at high energy colliders is studied by using renormalized pQCD. In section XI we discuss how hadron production at high energy colliders can be studied by using unrenormalized QCD. In section XII we discuss the advantages of the unrenormalized QCD over the renormalized QCD to study hadron production at high energy colliders. In section XIII we discuss that one can do renormalization in QCD to study hadron production at high energy colliders but it is not necessary because the same hadron production cross section can be calculated by using the unrenormalized QCD. Section XIV contains conclusions.

\section{ Infrared (Soft) Divergence and the Eikonal Feynman Rule}

It is well known that in quantum field theory the infrared divergence arises from the eikonal part of the Feynman diagram. We find that the study of the factorization of the infrared divergences due to the presence of light-like eikonal line is enormously simplified due to the pure gauge field. This can be seen as follows.

Let us first consider the QED before going to QCD as the eikonal Feynman rule in QCD is similar to that in QED. The contribution to the Feynman diagram for the process in which an incoming electron of four momentum $s^\mu$ emits a real photon of four momentum $k^\mu$ is given by \cite{gram}
\begin{eqnarray}
&& {\cal M}=\frac{1}{\gamma_\lambda s^\lambda -\gamma_\lambda k^\lambda -m} \gamma_\nu \epsilon^\nu(k)u(s)={\cal M}_{\rm eikonal}+{\cal M}_{\rm non-eikonal}
\label{at}
\end{eqnarray}
where
\begin{eqnarray}
{\cal M}_{\rm eikonal}=-\frac{s \cdot \epsilon(k)}{s \cdot k}u(s)
\label{aeik}
\end{eqnarray}
and
\begin{eqnarray}
{\cal M}_{\rm non-eikonal}=\frac{k^\lambda \gamma_\lambda \gamma_\nu \epsilon^\nu(k)}{2s \cdot k} u(s).
\label{anoneik}
\end{eqnarray}
Writing the photon field $\epsilon^\delta(k)$ as the sum of physical (transversely polarized) photon field $\epsilon^\delta_{\rm phys}(k)$ and the pure gauge (longitudinally polarized) photon field $\epsilon^\delta_{\rm pure}(k)$ we find
\begin{equation}
\epsilon_\delta(k) =\epsilon^{\rm phys}_\delta(k)+\epsilon^{\rm pure}_\delta(k)
\label{a}
\end{equation}
where \cite{gram}
\begin{equation}
\epsilon^{\rm phys}_\delta(k) = [\epsilon_\delta(k) -k_\delta \frac{s \cdot \epsilon(k)}{s \cdot k}]
\label{aa}
\end{equation}
and
\begin{equation}
\epsilon^{\rm pure}_\delta(k) = k_\delta \frac{s \cdot \epsilon(k)}{s \cdot k},
\label{ab}
\end{equation}
From eqs. (\ref{aeik}), (\ref{anoneik}), (\ref{a}), (\ref{aa}) and (\ref{ab}) we find in the infrared/soft photon limit $k_0,k_1,k_2,k_3 \rightarrow 0$ that
\begin{eqnarray}
{\cal M}^{\rm transverse~photon}_{\rm eikonal}=0, ~~~~~~~~~~~{\rm as}~~~~~~~~~~k_0,k_1,k_2,k_3 \rightarrow 0,
\label{tg}
\end{eqnarray}
\begin{eqnarray}
{\cal M}^{\rm longitudinal~photon}_{\rm non-eikonal}=0, ~~~~~~~~~~~{\rm as}~~~~~~~~~~k_0,k_1,k_2,k_3 \rightarrow 0,
\label{tf}
\end{eqnarray}
\begin{eqnarray}
{\cal M}^{\rm transverse~photon}_{\rm non-eikonal}  \rightarrow {\rm finite}, ~~~~~~~~~~~{\rm as}~~~~~~~~~~k_0,k_1,k_2,k_3 \rightarrow 0
\label{te}
\end{eqnarray}
and
\begin{eqnarray}
-{\cal M}^{\rm longitudinal~photon}_{\rm eikonal} \rightarrow \infty, ~~~~~~~~~~~{\rm as}~~~~~~~~~~k_0,k_1,k_2,k_3 \rightarrow 0.
\label{td}
\end{eqnarray}
From eqs. (\ref{tg}) and (\ref{te}) we find that in the infrared/soft limit the non-eikonal part of the Feynman diagram contributes to the finite (physical) cross section in quantum field theory but from eqs. (\ref{tf}) and (\ref{td}) we find that if the photon field is the pure gauge field (longitudinally polarized photon) then the infrared/soft divergence in quantum field theory can be studied by using the eikonal part of the Feynman diagram
without modifying the finite (physical) cross section.

Hence we find that if the photon field is the pure gauge field (longitudinally polarized photon) then the study of infrared divergence in quantum field theory can be simplified because the longitudinal polarization of the massless photon is un-physical which can be gauged away. In the next section we will show that the light-like eikonal line produces pure gauge field in quantum field theory.

\section{ Light-Like Eikonal Line in Quantum Field Theory generates Pure Gauge Field }

In the previous section we saw that pure gauge field (corresponding to longitudinally polarized photon) can simplify the study of infrared/soft divergence in quantum field theory. In this section we will show that light-like eikonal line produces pure gauge field in quantum field theory.

For the light-like electron of four-velocity $l^\mu$ we find from eq. (\ref{aeik}) that the eikonal contribution
\begin{eqnarray}
e\int \frac{d^4k}{(2\pi)^4} \frac{l \cdot Q(k)}{l \cdot k +i \epsilon}= i \int d^4x I(x) \cdot Q(x)
\label{eca}
\end{eqnarray}
gives the eikonal current density
\bea
I^\nu(x) =l^\nu ~e\int_0^\infty d\lambda \delta^{(4)}(x-l\lambda)
\label{ec}
\eea
where $Q^\delta(x)$ is the photon field.

In the path integral formulation of quantum field theory, the generating functional of the photon in the presence of external current density $I^\mu(x)$ is given by \cite{n1}
\bea
&& Z[I]=\int [dQ]
e^{i\int d^4x [-\frac{1}{4}[\partial^\nu Q^\lambda(x)-\partial^\lambda Q^\nu(x)][\partial_\nu Q_\lambda(x)-\partial_\lambda Q_\nu(x)] -\frac{1}{2 \alpha} (\partial_\nu Q^{\nu }(x))^2+ I(x) \cdot Q(x) ]}
\eea
which gives the normalized vacuum-to-vacuum transition amplitude $<0|0>_I$ in the presence of external source $I$
\bea
<0|0>_I=\frac{Z[I]}{Z[0]}= e^{i \int d^4x {\cal L}_{eff}(x)}
\label{zf}
\eea
where ${\cal L}_{eff}(x)$ is the effective lagrangian density.

Using eqs. (\ref{ec}) in (\ref{zf}) we find that the effective lagrangian density is given by \cite{n1}
\bea
{\cal L}_{eff}(x)=\frac{e^2(l^2)^2 }{2[(l \cdot x)^2]^2}=0, ~~~~~~~~l^2=0,~~~~~~~l\cdot x \neq 0
\label{ef}
\eea
which is the pure gauge field produced by the light-like ($l^2=0$) eikonal line at all time-space positions $x^\mu=(x_0,{\vec x})$ except at the spatial positions perpendicular to the motion of the charge (${\vec l} \cdot {\vec x} \neq 0$) at the time of closest approach $(t=x_0\neq 0)$.

Similarly the (interaction) effective lagrangian density in quantum field theory between the (light-like or non-light-like) non-eikonal current of four-momentum $l_1^\mu$ and the gauge field generated by the light-like eikonal current of four-velocity $l^\mu$ is given by \cite{n1}
\bea
{\cal L}^{int}_{eff}(x)=l^2 \frac{e^2}{2}\frac{(l \cdot l_1) (l_1 \cdot x) -(l \cdot x) l_1^2 }{(l \cdot x)^3[(l_1 \cdot x)^2 -l_1^2 x^2]^{\frac{3}{2}}}=0,~~~~~~{\rm for}~~~~~l_1\cdot x \neq 0,~~~~~~l\cdot x  \neq 0
\label{nf}
\eea
which is consistent with eq. (\ref{tf}).

Hence from eqs. (\ref{ef}) and (\ref{nf}) we find that the light-like eikonal current
produces pure gauge field in quantum field theory which is consistent with the corresponding result obtained in classical mechanics \cite{s1,n4,n5}.

In QED the U(1) pure gauge field $A^{\delta }(x)$ is given by $A_\delta(x) =\partial_\delta \omega(x)$ and in QCD the SU(3) pure gauge field $A^{\delta a}(x)$ is given by \cite{n1}
\bea
T^eA_\nu^e(x) =\frac{1}{ig} [\partial_\nu \Phi(x)]\Phi^{-1}(x),~~~~~~~~\Phi(x)={\cal P}e^{-igT^d\int_0^\infty d\lambda l\cdot A^d(x+l\lambda)}
\label{q8}
\eea
where $\Phi(x)$ is the light-like gauge link (or the eikonal line) in the fundamental representation of SU(3).

\section{ Collinear Divergence in Quantum Field Theory and the Pure Gauge Field }

Note from eqs. (\ref{aeik}) and (\ref{anoneik}) that for the non-light-like eikonal line the collinear divergence is absent in quantum field theory because for the non-light-like eikonal line $(s^2 \neq 0$) we have
\bea
s \cdot k \neq 0,~~~~~~~~~~{\rm when}~~~~~~~~~~~~s^2 \neq 0,~~~~~~~k^2=0,~~~~~k_0,k_1,k_2,k_3 \neq 0.
\label{lk}
\eea
From eqs. (\ref{aeik}) and (\ref{anoneik}) we find that the collinear divergence occurs when
\bea
s \cdot k=0,~~~~~~~~~{\rm or}~~~~~~~~~ l\cdot k=0,~~~~~~~~~~~{\rm for}~~~~~k_0,k_1,k_2,k_3 \neq 0
\label{cl}
\eea
which can only happen when the eikonal line is light-like ($s^2=0$ or $l^2=0$), {\it i. e.}, when
\bea
l^2=k^2=0.
\label{cl1}
\eea
But we have shown in section III that the light-like eikonal line ($l^2=0$) produces pure gauge field in quantum field theory. We have also shown that for the pure gauge field the non-eikonal contribution in eq. (\ref{anoneik}) vanishes, {\it i. e.},
\bea
{\cal M}^{\rm pure~gauge~field}_{\rm non-eikonal}=0
\eea
which is consistent with ${\cal L}^{int}_{eff}(x)=0$ in eq. (\ref{nf}).

Hence the collinear divergence arises from the eikonal part of the Feynman diagram given by eq. (\ref{aeik}). But as shown in eqs. (\ref{cl}) and (\ref{cl1}) one finds that the collinear divergence occurs when the eikonal line is light-like {\it i. e.}, $l \cdot k = 0$ when $l^2=0$. This implies that, since the light-like eikonal line produces pure gauge field, the collinear divergence due to the presence of light-like eikonal line in quantum field theory can be studied by using the pure gauge field without modifying the finite value of the (physical) cross section.

It is also useful to see how the pure-gauge property holds for the case in which the gauge field momentum is collinear to the momentum of the light-like eikonal line in quantum field theory. This can be seen as follows.

From the Feynman rule for the light-like eikonal line that is given in eq. (\ref{aeik}), it is apparent that, in the Feynman gauge the polarization of the gauge field is in the $l^\mu$ (or $s^\mu$) direction. The statement that the produced gauge field is a pure gauge is equivalent to the statement that the polarization of the gauge field is proportional to the momentum $k^\mu$ of the gauge field. For the gauge field produced by the light-like eikonal line ($l^2=0$), this is the case only if the momentum $k^\mu$ of the gauge field is collinear to $l^\mu$. This implies that the pure-gauge property holds for the case in which the gauge field momentum is collinear to the momentum of the light-like eikonal line.

Hence from the above discussions we find that the collinear divergence due to the presence of light-like eikonal line in quantum field theory can be studied by using the pure gauge field without modifying the finite value of the (physical) cross section.

\section{Relation Between N-Point Connected Green's Function in QCD and in the Background Field Method of QCD in the Presence of SU(3) Pure Gauge Background Field}

We have shown in the previous sections that the infrared/soft and collinear divergences in quantum field theory can be studied by using the pure gauge field. We have also shown in the previous sections that the light-like eikonal line produces pure gauge field both in quantum field theory and in classical mechanics. Hence in \cite{n1,n2,n3} we have proved the factorization of infrared/soft and collinear divergences in QCD by using the path integral formulation of the background field method of QCD in the presence of SU(3) pure gauge background field.

The renormalization in QCD by using the background field method of QCD is studied in \cite{ab}. What we study in this paper is that when the background field is the SU(3) pure gauge background field then the renormalization of ultra violet (UV) divergences and the factorization of infrared/soft and collinear divergences in QCD can be simultaneously studied at all orders in coupling constant by using the path integral formulation of the background field method of QCD.

In this section we will derive the relation between the connected green's function in QCD
and the connected green's function in the background field method of QCD in the presence of SU(3) pure gauge
background field.

\subsection{ N-Point Connected Green's Function in QCD }

In the path integral formulation the generating functional $Z[J,\eta,{\bar \eta}]$ in QCD is given by \cite{ab}
\bea
&&Z[J,\eta,{\bar \eta}]=\int [d{\bar \psi}][d\psi] [dQ] ~{\rm det}(\frac{\delta \partial^\delta Q_\delta^d}{\delta \omega^e})\nonumber \\
&&\times e^{i\int d^4x [-\frac{1}{4}F^{2}[Q] - \frac{1}{2\alpha} (\partial_\delta Q^{\delta e}(x))^2 +{\bar \psi}(x)[i\gamma^\delta \partial_\delta +gT^e \gamma^\delta Q_\delta^e(x) -m]\psi(x) + {\bar \psi}(x) \cdot \eta(x) +{\bar \eta}(x) \cdot \psi(x)+ J(x) \cdot Q(x)]}
\label{q1}
\eea
where $\alpha$ is the gauge fixing parameter and
\bea
F^2[Q]=F_{\delta \nu}^e[Q]F^{\delta \nu e}[Q]
\label{q2}
\eea
with
\bea
F_{\delta \nu}^e[Q] =\partial_\delta Q_\nu^e(x) -  \partial_\nu Q_\delta^e(x) + gf^{edc} Q_\delta^d(x) Q_\nu^c(x).
\label{q3}
\eea
In eq. (\ref{q1}) the term ${\rm det}( \frac{\delta \partial^\delta Q_\delta^d}{\delta \omega^e})$ is the ghost determinant which can be expressed in terms of the path integration over the ghost fields but we will directly work with the ghost determinant ${\rm det}(\frac{\delta \partial^\delta Q_\delta^d}{\delta \omega^e})$ in eq. (\ref{q1}) in this paper. Note that ${\bar \eta}_j(x)$ and $J_\delta^e(x)$ are the external sources for the quark field $\psi_j(x)$ and the quantum gluon field $Q^{\delta e}(x)$ respectively.

The connected Green's function of gluon in QCD is given by
\bea
G(x_1,...,x_n)=(-i)^{n-1}~ \frac{ \delta^n W[J,\eta,{\bar \eta}]}{\delta J(x_1)...\delta J(x_n)}|_{J=\eta ={\bar \eta} =0}
\label{r3}
\eea
where the suppression of Lorentz and color indices are understood where
\bea
W[J,\eta,{\bar \eta}]=\frac{1}{i}~{\rm ln}Z[J,\eta,{\bar \eta}].
\label{r5}
\eea

\subsection{ N-Point Connected Green's Function in the Background Field Method of QCD }

In the path integral formulation the generating functional $Z[A,J,\eta,{\bar \eta}]$ in the background field method of QCD is given by \cite{ab}
\bea
&&Z[A,J,\eta,{\bar \eta}]=\int [d{\bar \psi}][d\psi] [dQ] ~{\rm det}(\frac{\delta G^d(Q)}{\delta \omega^e})\nonumber \\
&&\times e^{i\int d^4x [-\frac{1}{4}F^{2}[Q+A] - \frac{1}{2\alpha} G^2(Q) +{\bar \psi}(x)[i\gamma^\delta \partial_\delta +gT^e \gamma^\delta (Q+A)_\delta^e(x) -m]\psi(x) + {\bar \psi}(x) \cdot \eta(x) +{\bar \eta}(x) \cdot \psi(x)+ J(x) \cdot Q(x)]} \nonumber \\
\label{q4}
\eea
where $A^{\delta e}(x)$ is the background field and
\bea
F^2[Q+A]=F_{\delta \nu}^e[Q+A]F^{\delta \nu e}[Q+A]
\label{q5}
\eea
with
\bea
F_{\delta \nu}^e[Q+A] =\partial_\delta [Q_\nu^e(x)+A_\nu^e(x)] -  \partial_\nu [Q_\delta^e(x)+A_\delta^e(x)] + gf^{edc} [Q_\delta^d (x)+A_\delta^d(x)] [Q_\nu^c(x)+A_\nu^c(x)].\nonumber \\
\label{q6}
\eea
In eq. (\ref{q4}) the gauge fixing term
\bea
G^e(Q) = \partial^\delta Q_\delta^e(x) + gf^{edc} A_\delta^d(x) Q^{\delta c}(x)=D^\delta[A]Q_\delta^e(x)
\label{q7}
\eea
depends on the background field $A^{\delta e}(x)$. The term $~{\rm det}(\frac{\delta G^d(Q)}{\delta \omega^e})$ is the ghost determinant which can be expressed in terms of the path integration over the ghost fields but we will directly work with the ghost determinant $~{\rm det}(\frac{\delta G^d(Q)}{\delta \omega^e})$ in eq. (\ref{q4}) in this paper.

Analogous to $W[J,\eta,{\bar \eta}]$ in eq. (\ref{r5}) we have \cite{ab}
\bea
W[A,J,\eta,{\bar \eta}]=\frac{1}{i}~{\rm ln}Z[A,J,\eta,{\bar \eta}]
\label{br5}
\eea
in the background field method of QCD. The connected Green's function of gluon in the background field method of QCD is given by
\bea
G^A(x_1,...,x_n)=(-i)^{n-1}~ \frac{ \delta^n W[A,J,\eta,{\bar \eta}]}{\delta J(x_1)...\delta J(x_n)}|_{J=\eta ={\bar \eta} =0}
\label{br3}
\eea
where the suppression of Lorentz and color indices are understood.

\subsection{ Relation Between N-Point Connected Green's Function in QCD and in the Background Field Method of QCD in the Presence of SU(3) Pure Gauge Background Field}

When the background field $A^{\delta e}(x)$ is the SU(3) pure gauge background field given by eq. (\ref{q8})
we find \cite{n1,n2,n3}
\bea
Z[A,J',\eta',{\bar \eta}']=Z[J,\eta,{\bar \eta}],~~~~~~~~~~~~~~~~~~~~W[A,J',\eta',{\bar \eta}']=W[J,\eta,{\bar \eta}]
\label{q9}
\eea
where
\bea
\eta'(x) =\Phi(x)\eta(x),~~~~~~~~~J'(x)=\Phi^{(A)}(x)~J(x)
\label{q10}
\eea
where $\Phi(x)$ is the light-like gauge link in the fundamental representation of SU(3) as given by eq. (\ref{q8}) and
\bea
\Phi^{(A)}(x)={\cal P}e^{-igT^{(A)d}\int_0^\infty d\lambda l\cdot A^d(x+l\lambda)},~~~~~~~~T^{(A)d}_{ec}=-if^{dec}
\label{a40}
\eea
is the light-like gauge link in the adjoint representation of SU(3).

Hence from eqs. (\ref{q9}), (\ref{r3}) and (\ref{br3}) we find
\bea
&& G^A(x_1,...,x_n)=\Phi^{(A) }(x_1)...\Phi^{(A) }(x_n)G(x_1,...,x_n)
\label{a38}
\eea
where $G(x_1,...,x_n)$ is the n-point connected green's function of gluon in QCD and
$G^A(x_1,...,x_n)$ is the n-point connected green's function of gluon in QCD in
the presence of SU(3) pure gauge background field $A^{\delta e}(x)$ where $\Phi^{(A)}(x)$ is the light-like gauge-link (or the eikonal line) in the adjoint representation of SU(3) as given by eq. (\ref{a40}).

\subsection{ Relation Between (Full) Propagator in QCD and in the Background Field Method of QCD in the Presence of SU(3) Pure Gauge Background Field}

From eq. (\ref{q9}) we find
\bea
&& G^A(x_1,x_2)=\Phi^{(A)}(x_1)\Phi^{(A)}(x_2)G(x_1,x_2)
\label{a41x}
\eea
where $G(x_1,x_2)$ is the (full) propagator of gluon in QCD and
$G^A(x_1,x_2)$ is the (full) propagator of gluon in QCD in
the presence of SU(3) pure gauge background field $A^{\mu a}(x)$ where $\Phi^{(A)}(x)$
is the light-like gauge-link (or the eikonal line) in the adjoint representation of SU(3) as given by eq. (\ref{a40}).

\section{ Renormalization of the Wilson Line }

In the background field method of QCD the relation between the unrenormalized and renormalized background field and coupling
constant are given by \cite{ab}
\bea
&& A^{\mu a}=Z^{\frac{1}{2}}_A A^{\mu a}_R, \nonumber \\
&& g = Z_g g_R
\label{a30}
\eea
where $Z_A,Z_g$ are divergent constants and the subscript $R$ stands for renormalized quantities.

The $\beta$ function in QCD can be obtained from the renormalization of the background field $A_\mu^a$ from the condition \cite{ab}
\bea
gA=g_R A_R.
\label{a41}
\eea
Hence by using eq. (\ref{a41}) in (\ref{a40}) we find
\bea
\Phi^{(A)}(x)=\Phi^{(A)}_R(x)
\label{a42}
\eea
which proves that unrenormalized Wilson line is same as renormalized Wilson line in the background field method of QCD.

\section{ S-Matrix Element in QCD at all orders in coupling constant and Parton Distribution/Fragmentation Function at high energy colliders}

Consider all the external particles as gluons in the general scattering process in QCD
\bea
p_1+p_2 \rightarrow p'_1+p'_2+...+p'_n.
\label{h37mg}
\eea
Using LSZ reduction formula the S-matrix element for the general scattering process
in eq. (\ref{h37mg}) in QCD at all orders in coupling constant is given by \cite{nlsz}
\bea
&&<f|i> =\int d^4x'_1... \int d^4x'_n \int d^4x_2  \int d^4x_1 ~e^{i p'_1 \cdot x'_1+...+i p'_n \cdot x'_n-i p_2 \cdot x_2-i p_1 \cdot x_1} \int d^4y'_1...\int d^4y'_n \int d^4y_2 \int d^4y_1 \nonumber \\
&& \times [G_R(x'_1,y'_1)]^{-1}... [G_R(x'_n,y'_n)]^{-1}[G_R(x_2,y_2)]^{-1}[G_R(x_1,y_1)]^{-1}~G_R(y'_1,...,y'_n,y_2,y_1)
\label{nbg}
\eea
where $G_R(x_1,x_2)$ is the renormalized (full) propagator of gluon and $G_R(x_1,x_2,...,x_n)$ is the renormalized n-point connected green's function of gluon. The suppression of color and Lorentz indices in $G_R(x_1,x_2)$ and $G_R(x_1,x_2,...,x_n)$ are understood.

The partonic level scattering cross section ${\hat \sigma}$ in QCD at all orders in coupling constant is proportional to the S-matrix element square in QCD
\bea
{\hat \sigma} \propto |{\cal M}|^2,~~~~~~~~~~|{\cal M}|^2 \propto |<f|i>|^2.
\label{ps}
\eea
Note that since we are interested in the ultra violet (UV), infrared (IR) and collinear divergences in QCD in this paper, the finite factors due to the relevant sum of polarization vectors (and color factors) are included in the partonic level cross section ${\hat \sigma}$ in eq. (\ref{ps}) instead of the S-matrix element in eq. (\ref{nbg}) in order to simplify the calculation.

Extending eq. (\ref{nbg}) we find that the S-matrix element in QCD in the presence of SU(3) pure gauge background field $A^{\mu a}(x)$ is given by
\bea
&&<f|i>_A =\int d^4x'_1... \int d^4x'_n \int d^4x_2  \int d^4x_1 ~e^{i p'_1 \cdot x'_1+...+i p'_n \cdot x'_n-i p_2 \cdot x_2-i p_1 \cdot x_1} \int d^4y'_1...\int d^4y'_n \int d^4y_2 \int d^4y_1 \nonumber \\
&& \times [G^A_R(x'_1,y'_1)]^{-1}... [G^A_R(x'_n,y'_n)]^{-1}[G^A_R(x_2,y_2)]^{-1}[G^A_R(x_1,y_1)]^{-1}~G^A_R(y'_1,...,y'_n,y_2,y_1)
\label{bgf}
\eea
where $G^A_R(x_1,x_2)$ is the renormalized (full) propagator of gluon in the presence of SU(3) pure gauge background field $A^{\mu a}(x)$ and $G^A_R(x_1,x_2,...,x_n)$ is the renormalized n-point connected green's function of gluon in the presence of SU(3) pure gauge background field $A^{\mu a}(x)$.

Folding the gluon distribution function $f$ inside hadron and the gluon to hadron fragmentation function $D$ with the partonic level scattering cross section ${\hat \sigma}$ in QCD in eq. (\ref{ps}) we find that the hadronic level cross section $\sigma$ for the general scattering process in eq. (\ref{h37mg}) at high energy colliders is given by
\bea
{\sigma} = f_1 \otimes f_2 \otimes {\hat \sigma} \otimes D_1 \otimes ....\otimes D_n.
\label{ps2}
\eea

The unrenormalized gluon distribution function $f^{UnRenormalized}$ inside the hadron $H$ or the unrenormalized gluon to hadron fragmentation function $D^{UnRenormalized}$ is proportional to the gauge invariant 2-point (quantum) gluon correlation function of the type \cite{n3}
\bea
&& f_i^{UnRenormalized} = B \int d^3x_i e^{i{\bf k}_i \cdot {\bf x}_i} <H|\Phi^{(A)\dagger}({\bf x}_i)Q_\mu^{ a}({\bf x}_i)\Phi^{(A)\dagger}(0)Q^{\mu a}(0)|H>,\nonumber \\
&&D_i^{UnRenormalized} =  C \int d^3x_i e^{i{\bf k}_i \cdot {\bf x}_i} <0|\Phi^{(A)\dagger}({\bf x}_i)Q_\mu^{ a}({\bf x}_i)a^\dagger_H a_H \Phi^{(A)\dagger}(0)Q^{\mu a}(0)|0>
\label{ps1}
\eea
where $B,C$ are some finite factors, $a^\dagger_H$ is the creation operator of the hadron $H$ and $\Phi^{(A)}(x)$ is the gauge-link (or the eikonal line) in the adjoint representation of SU(3) as given by eq. (\ref{a40}).

Similarly the renormalized gluon distribution function $f$ inside the hadron or
the renormalized gluon to hadron fragmentation function $D$ is given by
\bea
&& f_i=B \int d^3x_i e^{i{\bf k}_i \cdot {\bf x}_i} <H|\Phi^{(A)\dagger}_R({\bf x}_i)Q_{\mu R}^a({\bf x}_i)\Phi^{(A)\dagger}_R(0)Q^{\mu a}_R(0)|H>,\nonumber \\
&&D_i =C \int d^3x_i e^{i{\bf k}_i \cdot {\bf x}_i}<0|\Phi^{(A)\dagger}_R({\bf x}_i)Q_{\mu R}^a({\bf x}_i) a^\dagger_H a_H  \Phi^{(A)\dagger}_R(0)Q^{\mu a}_R(0)|0>.
\label{ps1a}
\eea

\section{ Proof of Factorization in QCD at all orders in coupling constant and Cancelation of Infrared and Collinear Divergences in Hadrons Production at High Energy Colliders}

Using eqs. (\ref{a38}), (\ref{a41x}) and (\ref{a42}) in eq. (\ref{bgf}) we find
\bea
&&<f|i>_A =\int d^4x'_1... \int d^4x'_n \int d^4x_2  \int d^4x_1 e^{i p'_1 \cdot x'_1+...+i p'_n \cdot x'_n-i p_2 \cdot x_2-i p_1 \cdot x_1} \nonumber \\
&&\times  \Phi^{(A)\dagger }(x_1)\Phi^{(A)\dagger }(x_2)\Phi^{(A)\dagger}(x'_1)...\Phi^{(A)\dagger}(x'_n) \int d^4y'_1...\int d^4y'_n \int d^4y_2 \int d^4y_1 \nonumber \\
&& \times [G_R(x'_1,y'_1)]^{-1}... [G_R(x'_n,y'_n)]^{-1}[G_R(x_2,y_2)]^{-1}[G_R(x_1,y_1)]^{-1}~G_R(y'_1,...,y'_n,y_2,y_1).
\label{bgff}
\eea
where we have used
\bea
&& Q^{\mu a}=Z_Q Q^{\mu a}_R
\label{qa30}
\eea
where $Z_Q$ is the divergent constant from the (quantum) gluon field $Q^{\mu a}$ renormalization.

Eqs. (\ref{nbg}) and (\ref{bgff}) prove that the infrared and collinear divergences due to the presence of light-like eikonal line are factorized into the gauge links $\Phi^{(A)}(x)$.

Similarly by using eqs. (\ref{a41x}), (\ref{a38}), (\ref{qa30}) and (\ref{a42}) in eq. (\ref{nbg}) we find that the S-matrix element in QCD is given by
\bea
&&<f|i> = \int d^4x'_1... \int d^4x'_n \int d^4x_2  \int d^4x_1 ~e^{i p'_1 \cdot x'_1+...+i p'_n \cdot x'_n-i p_2 \cdot x_2-i p_1 \cdot x_1} \nonumber \\
&& \times \Phi^{(A) }_R(x'_1)...\Phi^{(A) }_R(x'_n) \Phi^{(A) }_R(x_2)\Phi^{(A)}_R(x_1)\int d^4y'_1...\int d^4y'_n \int d^4y_2 \int d^4y_1 \nonumber \\
&& \times [G^A_R(x'_1,y'_1)]^{-1}... [G^A_R(x'_n,y'_n)]^{-1}[G^A_R(x_2,y_2)]^{-1}[G^A_R(x_1,y_1)]^{-1}~G^A_R(y'_1,...,y'_n,y_2,y_1).
\label{nbgf}
\eea
Similar to above the eqs. (\ref{bgf}) and (\ref{nbgf}) prove that the infrared and collinear divergences due to the presence of light-like eikonal line are factorized into the gauge links $\Phi^{(A)}(x)$.

For the scattering process in eq. (\ref{h37mg}) the S-matrix element in QCD at all orders in coupling constant is given by eq. (\ref{nbgf}) and the S-matrix in QCD in the presence of SU(3) pure gauge background field $A^{\delta e}(x)$ is given by eq. (\ref{bgf}). The only difference between eqs. (\ref{bgf}) and (\ref{nbgf}) are the gauge links $\Phi^{(A)}(x)$. Hence by comparing eq. (\ref{bgf}) with (\ref{nbgf}) we find that when the gauge links are supplied at each external particles in the S-matrix element in QCD in the presence of SU(3) pure gauge background field $A^{\delta e}(x)$ one reproduces the S-matrix element in QCD at all orders in coupling constant.

In order to see how the cancelation of infrared/soft and collinear divergences happen in the hadrons production at high energy colliders it is easy to start with the definition of the PDF and FF in eq. (\ref{ps1a}) which are obtained from the proof of the factorization of the 2-point non-perturbative correlation function in QCD given by \cite{n1,n2,n3}
\bea
<Q_\delta^e(x_1) Q_\nu^c(x_2)>=<\Phi^{(A)\dagger}(x_1)Q_\delta^e(x_1) \Phi^{(A)\dagger}(x_2)Q_\nu^c(x_2)>_A.
\label{fb}
\eea
In eq. (\ref{fb}) the $<Q_\delta^e(x_1) Q_\nu^c(x_2)>$ in the left hand side is the 2-point non-perturbative correlation function in QCD and the $<Q_\delta^e(x_1) Q_\nu^c(x_2)>_A$ in the right hand side is the 2-point non-perturbative correlation function in the background field method of QCD in the presence of SU(3) pure gauge background field $A^{\delta e}(x)$. Note that the eq. (\ref{fb}) can also be directly obtained from eq. (\ref{q9}). Hence in the definition of the PDF and FF the 2-point non-perturbative correlation functions in eq. (\ref{ps1a}) are in the presence of the light-like eikonal line. Similarly the green's functions in the right hand side of eq. (\ref{nbgf}) are in the presence of light-like eikonal line. Hence one finds that the infrared/soft and collinear divergences contained in the light-like gauge links in eq. (\ref{nbgf}) exactly cancel with the corresponding infrared/soft and collinear divergences contained in the light-like gauge links in eq. (\ref{ps1a}) when used in the hadrons production cross section formula in eq. (\ref{ps2}).

Another easy way to see this is to reverse the eq. (\ref{fb}) to find
\bea
<Q_\delta^e(x_1) Q_\nu^c(x_2)>_A=<\Phi^{(A)}(x_1)Q_\delta^e(x_1) \Phi^{(A)}(x_2)Q_\nu^c(x_2)>
\label{fc}
\eea
where the right hand side is in QCD and the left hand side is in the background field method of QCD in the presence of SU(3) pure gauge background field $A^{\mu a}(x)$. Hence eqs. (\ref{bgff}) and (\ref{fc}) prove the cancelation of infrared and collinear divergences contained in the light-like gauge links in a similar way described above.

Note that eqs. (\ref{ps1a}) and (\ref{nbgf}) correspond to the scattering process in eq. (\ref{h37mg}) for which the proof of cancelation of infrared/soft and collinear divergences in the hadrons production at high energy colliders is given above and below eq. (\ref{fb}).

\section{ Simultaneous Proof of Renormalization and Factorization in QCD at All Orders in Coupling Constant at High Energy Coliders }

From eqs. (\ref{qa30}), (\ref{a42}) and (\ref{nbgf}) we find
\bea
&&<f|i> =Z_Q^{(n+2)} <f|i>^{UnRenormalized}
\label{smb}
\eea
where the unrenormalized S-matrix element $<f|i>^{UnRenormalized}$ in QCD at all orders in coupling constant is given by
\bea
&&<f|i>^{UnRenormalized}=\int d^4x'_1\int d^4x'_2... \int d^4x'_n \int d^4x_2  \int d^4x_1 ~e^{i p'_1 \cdot x'_1+i p'_2 \cdot x'_2+...+i p'_n \cdot x'_n-i p_2 \cdot x_2-i p_1 \cdot x_1} \nonumber \\
&&\int d^4y'_1\int d^4y'_2...\int d^4y'_n  \int d^4y_2 \int d^4y_1 ~ [G(x'_1,y'_1)]^{-1}G(x'_2,y'_2)]^{-1}... [G(x'_n,y'_n)]^{-1}\nonumber \\
&&[G(x_2,y_2)]^{-1}[G(x_1,y_1)]^{-1}~G(y'_1,y'_2,...,y'_n,y_2,y_1)
\label{bnbgf}
\eea
where $G(x_1,x_2)$ is the unrenormalized (full) propagator of gluon and $G(x_1,x_2,...,x_n)$ is the unrenormalized n-point connected green's function of gluon.

Note that eq. (\ref{smb}) can also be directly obtained from eqs. (\ref{qa30}) and (\ref{nbg}).

From eqs. (\ref{a42}), (\ref{qa30}), (\ref{ps1}) and (\ref{ps1a}) we find
\bea
&& f_i = Z_Q^{-2} ~f_i^{UnRenormalized}, \nonumber \\
&& D_i = Z_Q^{-2} ~D_i^{UnRenormalized}.
\label{ps1b}
\eea
From eqs. (\ref{smb}), (\ref{ps1b}), (\ref{ps}) and (\ref{ps2}) we find
\bea
&&\sigma=f_1 \otimes f_2 \otimes {\hat \sigma} \otimes D_1 \otimes ....\otimes D_n=\nonumber \\
&&f_1^{UnRenormalized} \otimes f_2^{UnRenormalized} \otimes {\hat \sigma}^{UnRenormalized} \otimes D_1^{UnRenormalized} \otimes ...\otimes D_n^{UnRenormalized}\nonumber \\
\label{final}
\eea
which reproduces eq. (\ref{sgu}) which proves that the renormalization in QCD is not necessary to study hadrons production at high energy colliders at all orders in coupling constant.

From eqs. (\ref{nbgf}), (\ref{ps1a}), (\ref{final}), (\ref{ps}) and (\ref{ps2}) we find that the path integral formulation of the background field method of QCD in the presence of SU(3) pure gauge background field enormously simplifies the simultaneous proof of the renormalization of ultra violet (UV) divergences and the factorization of infrared (IR) and collinear divergences in QCD at all orders in coupling constant at high energy colliders.

\section{ Present Approach of Renormalized Perturbative QCD to Study Hadron Production at High Energy Colliders }

In this section we will briefly review the present approach of renormalized pQCD which has been widely used to study hadron production at high energy colliders. In the next section we will describe that the renormalization in QCD is not necessary to study hadron production cross section at high energy colliders where we will discuss how one can study hadron production cross section at high energy colliders in unrenormalized QCD.

\subsection{Renormalized QCD Coupling is Not a Physical Observable}\label{gr}

Within the perturbative QCD (pQCD) one calculates the finite partonic level cross section ${\hat \sigma}$ at LO, NLO, NNLO etc. in coupling constant by renormalizing QCD because in the unrenormalized QCD the partonic level cross section ${\hat \sigma}^{UnRenormalized}$ can become infinite due to the ultra violet (UV) divergences in the loop diagrams. The dimensional regularization in the $4-2\epsilon$ dimensions is commonly used in the literature where the $\epsilon$ serves as a regulator to separate the finite part from the UV divergent part that appears as $\frac{1}{\epsilon^n}$ poles in the limit $\epsilon \rightarrow 0$. If one holds the bare coupling constant $g$ fixed then one gets the divergent partonic cross section in the limit $\epsilon \rightarrow 0$ in the unrenormalized QCD, see subsection \ref{gur}. In order to calculate the finite partonic cross section in the renormalized QCD one adjusts the bare coupling constant $g$ as follows before setting $\epsilon \rightarrow 0$.

One writes the bare coupling constant $g$ in terms of renormalized running coupling $g_R(\mu)$ via the equation
\bea
g=g_R(\mu) Z_g \mu^{2\epsilon}
\label{bc}
\eea
where $\mu$ is the unphysical mass scale and $Z_g$ is the divergent renormalization constant in the limit $\epsilon \rightarrow 0$. Since the bare coupling constant $g$ is independent of the unphysical mass scale $\mu$ one finds that the renormalized running coupling $g_R(\mu)$ depends on the unphysical mass scale $\mu$. This is the origin of the mass scale $\mu$ in renormalized pQCD.

After adding the counter terms to remove ultra violet (UV) divergences from the loop integrals the mass scale $\mu$ still remains in the finite part of the loop integral. For example one can set the renormalized running coupling at the scale of the Z boson mass to \cite{alp}
\bea
\alpha_s(M_Z)=\frac{g^2_R(M_Z)}{4\pi}=0.118.
\eea
It should be mentioned here that the mass scale $\mu$ in QCD is unphysical. Since a physical observable in QCD can not depend on the unphysical mass scale $\mu$ one finds that the running coupling $g_R(\mu)$ in renormalized QCD is not a physical observable.

\subsection{Finite Partonic Cross Section in Renormalized QCD is Not Physical Observable }\label{pcs}

After renormalization the finite partonic cross section ${\hat \sigma}$ in the renormalized pQCD calculation is expressed in terms of the renormalized coupling $g_R(\mu)$. However, from the discussion from the subsection \ref{gr} we saw that the renormalized coupling $g_R(\mu)$ depends on the unphysical mass scale $\mu$. Since the renormalized finite partonic cross section ${\hat \sigma}$ in the renormalized pQCD depends on the unphysical mass scale $\mu$ one finds that the renormalized finite partonic cross section ${\hat \sigma}$ in the renormalized pQCD is not a physical observable. This is obvious because the parton is not a physical observable due to confinement in QCD. As mentioned in subsection \ref{gr} since the partonic cross section ${\hat \sigma}$ is not a physical observable it is not necessary to make it finite which means it is not necessary to renormalize QCD to make the partonic cross section finite, see also subsection \ref{uin}. All that matters is that the hadronic cross section (which is a physical observable) is finite in QCD which we have discussed in subsections \ref{rge} and \ref{fur}.

\subsection{  Finite Parton Distribution/Fragmentation Function in Renormalized QCD is Not Physical Observable }

Similarly after renormalization the finite parton distribution function (PDF) $f_i$ inside the hadron and the finite parton to hadron fragmentation function (FF) $D_i$ depend on the unphysical mass scale $\mu$. Since the $f_i$ and $D_i$ in renormalized QCD  depend on the unphysical mass scale $\mu$ we find that the PDF and FF in renormalized QCD are not physical observable. This is similar to the partonic cross section ${\hat \sigma}$ which is not a physical observable in renormalized QCD which we discussed in subsection \ref{pcs}. Since the PDF $f_i$ and the FF $D_i$ are not physical observable there is no need to make them finite and hence it is not necessary to renormalize QCD to make the PDF $f_i$ and the FF $D_i$ finite. This is similar to the situation of the partonic cross section ${\hat \sigma}$ which is discussed in the subsection \ref{pcs}, see also the discussion in the subsection \ref{uina}.

It should be mentioned here that for very small values of the renormalized coupling $g_R(\mu)$ the finite partonic cross section ${\hat \sigma}$ in the renormalized QCD is usually calculated by using the pQCD, for example at LO, NLO, NLLO etc. in the coupling constant. However, the renormalized pQCD can not calculate the finite parton distribution function (PDF) $f_i$ inside the hadron and the finite parton to hadron fragmentation function (FF) $D_i$ because they are non-perturbative quantities in QCD. Because of this reason the finite value of the parton distribution function (PDF) $f_i$ inside the hadron and the finite value of the parton to hadron fragmentation function (FF) $D_i$ in the renormalized pQCD are extracted from the experiments.

\subsection{ Finite Hadron Cross Section in Renormalized QCD at High Energy Colliders is Physical Observable}\label{rge}

The physical observable in QCD is the hadron. Hence the hadron production cross section in renormalized pQCD at high energy colliders should not depend on the unphysical mass scale $\mu$. In order to make sure that the hadron production cross section in pQCD is independent of the unphysical mass scale $\mu$ one is forced to depend on additional constraint equations which are known as the renormalization group equations, see subsection \ref{rgn}. After the factorization of infrared (IR) and collinear divergences is proved, the finite value of the hadron production cross section $\sigma$ at high energy colliders is obtained from the finite partonic cross section ${\hat \sigma}$ by convoluting it with the finite PDF $f_i$ and the finite FF $D_i$ in eq. (\ref{ps2}) in the renormalized pQCD. This is the usual procedure in the renormalized pQCD to study hadron production at high energy colliders.

Note that, as mentioned above, the renormalization group equations are the additional constraints in renormalized pQCD because the renormalization group equations are necessary in renormalized pQCD to make sure that the physical observable (the hadronic cross section) is independent of the unphysical mass scale $\mu$, see subsection \ref{rgn} for more details. In case of unrenormalized QCD no such the renormalization group equations are necessary to study physical observable (the hadronic cross section) at high energy colliders, see subsection \ref{rgun}. Hence the unrenormalized QCD has advantage over the renormalized QCD in this respect which is one among several advantages listed in section \ref{adv}.

\section{ Hadron Production at High Energy Colliders Can be Studied by Using Unrenormalized QCD }

In this section we will discuss how one can study finite hadron production cross section at high energy colliders in unrenormalized QCD. In the next section we will describe the advantage of the unrenormalized QCD over the renormalized QCD to study hadron production at high energy colliders.

First of all we note that our final result in eq. (\ref{sgu}) is an exact result in QCD. This is because we have used the path integral formulation of QCD to derive it. The LSZ reduction formula for the partonic scattering in eq. (\ref{nbg}) is the exact formula in QCD because it uses the (full) connected Green's function and the (full) propagator from eq. (\ref{r3}) which are obtained from the generating functional from eq. (\ref{q1}) by using the path integral formulation of QCD.
Similarly the ultra violet (UV), infrared (IR) and collinear divergences behavior of the non-perturbative parton distribution function (PDF) inside the hadron and the non-perturbative parton to hadron fragmentation function (FF) in eqs. (\ref{ps1a}) and (\ref{ps1b}) are studied by using the path integral formulation of QCD \cite{n3}. Hence our final result in eq. (\ref{sgu}) is an exact result in QCD which is valid at all orders in coupling constant and is valid for any values of the QCD coupling constant.

In contrast to this the perturbative QCD (pQCD) is applicable only for very small values of the QCD coupling constant. Hence the perturbative QCD can not always predict an exact result in QCD. For example, the parton distribution function (PDF) and the parton to hadron fragmentation function (FF) in eq. (\ref{ps1a}) can not be studied by using the perturbative QCD because they are non-perturbative quantities in QCD. Take for example a non-perturbative function of the type
\bea
f(g) =e^{-\frac{1}{g^n}},~~~~~~~~~~n\ge 2.
\label{np}
\eea
The Taylor series at $g = 0$ for this function $f(g)$ is exactly zero to all orders in perturbation theory, but the function $f(g)$ is non-zero if $g \neq 0$. Hence one finds that the properties of the non-perturbative quantities in QCD like PDF and FF may not be correctly studied by using pQCD no matter how many orders of perturbation theory is used. On the other hand the path integral formulation of QCD which we have used in this paper can correctly predict the properties of the non-perturbative quantities in QCD.

\subsection{ Bare Coupling Constant and Bare Quark Mass in Unrenormalized QCD  }\label{gur}

Note that in order to carry out renormalization by adding counter terms, the bare coupling constant $g$ in eq. (\ref{bc}) in four dimensions is assumed to be infinite in renormalized QCD. The argument given for this is that the bare coupling constant $g$ in the renormalized QCD is not physical. But as we saw in subsection \ref{gr} the renormalized coupling $g_R(\mu)$ in renormalized QCD is not a physical observable because it depends on the unphysical mass scale $\mu$. Similarly the renormalized running quark mass $m_R(\mu)$ in renormalized QCD is not a physical observable because it depends on the unphysical mass scale $\mu$. The physical observable in QCD is the hadronic cross section which is independent of the unphysical mass scale $\mu$.

Hence we can keep the bare coupling constant $g$ and bare quark mass $m$ and obtain an infinite partonic cross section in unrenormalized QCD. There is nothing wrong if the partonic cross section in unrenormalized QCD becomes infinite because the partonic cross section is not physical observable. What matters is that the hadron production cross section (which is the physical observable) is finite at high energy colliders in unrenormalized QCD which we have discussed in subsection \ref{fur}.

\subsection{ Infinite Partonic Cross Section in Unrenormalized QCD is Not Physical Observable }\label{uin}

As mentioned in subsection \ref{gur} we can keep the bare coupling constant $g$ and the bare quark mass $m$ and obtain an infinite partonic cross section ${\hat \sigma}^{UnRenormalized}$ in unrenormalized QCD. Since the partonic cross section is not physical observable there is nothing wrong if the partonic cross section ${\hat \sigma}^{UnRenormalized}$ in unrenormalized QCD becomes infinite. The important thing is to separate the finite part of the partonic cross section from the UV divergent part of the partonic cross section in the unrenormalized QCD. For example, if one uses the dimensional regularization in the $4-2\epsilon$ dimensions then one can separate the finite part of the partonic cross section from the UV divergent part of the partonic cross section in the limit $\epsilon \rightarrow 0$ in the unrenormalized QCD. The UV divergent part of the partonic cross section in the unrenormalized QCD appears as the $\frac{1}{\epsilon^n}$ poles in the limit $\epsilon \rightarrow 0$ which will cancel with the corresponding $\frac{1}{\epsilon^n}$ poles in the limit $\epsilon \rightarrow 0$ from the parton distribution functions (PDFs) and fragmentation functions (FFs) in the unrenormalized QCD, see subsection \ref{hur}. The finite part of the partonic cross section in unrenormalized QCD becomes function of the bare coupling constant $g$ and the bare quark mass $m$. This finite part of the partonic cross section in unrenormalized QCD contributes to the finite hadron production cross section at high energy colliders in unrenormalized QCD, see subsection \ref{fur}.

\subsection{ Infinite Parton Distribution/Fragmentation Function in Unrenormalized QCD is Not Physical Observable }\label{uina}

Similarly as mentioned in subsection \ref{uin} we can keep the bare coupling constant $g$ and the bare quark mass $m$ and obtain infinite parton distribution function (PDF) ${f_i}^{UnRenormalized}$ and the infinite fragmentation function (FF) ${D_i}^{UnRenormalized}$ in unrenormalized QCD. Since the parton distribution function (PDF) and the fragmentation function (FF) are not physical observable there is nothing wrong if they become infinite in unrenormalized QCD. The important thing is to separate the finite parts from the UV divergent parts. For example, similar to the partonic cross section analysis in subsection \ref{uin}, if one uses the dimensional regularization procedure the UV divergent parts of the PDFs and FFs in the unrenormalized QCD appear as the $\frac{1}{\epsilon^n}$ poles in the limit $\epsilon \rightarrow 0$ which will cancel with the corresponding $\frac{1}{\epsilon^n}$ poles in the limit $\epsilon \rightarrow 0$ from the partonic cross section in the unrenormalized QCD. The finite parts of the PDFs and FFs in the unrenormalized QCD depend on the bare coupling constant $g$ and the bare quark mass $m$. The finite parts of the PDFs and FFs in unrenormalized QCD contribute to the finite hadron production cross section at high energy colliders in unrenormalized QCD, see subsection \ref{fur}.

\subsection{ Finite Hadron Cross Section in Unrenormalized QCD at High Energy Colliders is Physical Observable}\label{fur}

We have proved in eq. (\ref{sgu}) that the finite hadron production cross section $\sigma$ at high energy colliders in unrenormalized QCD is given by
\bea
\sigma=f_1^{UnRenormalized} \otimes f_2^{UnRenormalized} \otimes {\hat \sigma}^{UnRenormalized} \otimes D_1^{UnRenormalized} \otimes ...\otimes D_n^{UnRenormalized}\nonumber \\
\label{fint}
\eea
which is exactly the same finite hadron production cross section $\sigma$ at high energy colliders in renormalized QCD given by
\bea
\sigma=f_1^{Renormalized} \otimes f_2^{Renormalized} \otimes {\hat \sigma}^{Renormalized} \otimes D_1^{Renormalized} \otimes ...\otimes D_n^{Renormalized}.\nonumber \\
\label{fintr}
\eea

Eq. (\ref{sgu}) proves that the ultra violet (UV) divergences in the unrenormalized partonic level cross section ${\hat \sigma}^{UnRenormalized}$ exactly cancel with the ultra violet (UV) divergences in the unrenormalized parton distribution functions $f_i^{UnRenormalized}$ and with the UV divergences in the unrenormalized parton fragmentation functions $D_i^{UnRenormalized}$at all orders in coupling constant.

Hence from eq. (\ref{sgu}) one finds that the finite cross section of the hadron production at high energy colliders can be studied in the unrenormalized QCD. In addition to this, as discussed in subsections \ref{uin} and \ref{uina}, one finds that the finite hadronic cross section at high energy colliders in unrenormalized QCD is function of the bare coupling $g$ and the bare quark mass $m$. Hence the unrenormalized QCD has lot of advantages over the renormalized QCD to study hadron production at high energy colliders which we will discuss in section \ref{adv}.

\subsection{Cancelation of UV Divergences in Hadron Production at High Energy Colliders in Unrenormalized QCD is Similar to Cancelation of IR and Collinear Divergences}\label{hur}

As discussed in subsection \ref{uin} if one uses the dimensional regularization in $4-2\epsilon$ dimensions then the UV divergent part of the partonic cross section  ${\hat \sigma}^{UnRenormalized}$ in the unrenormalized QCD appears as the $\frac{1}{\epsilon^n}$ poles in the limit $\epsilon \rightarrow 0$. Similarly as discussed in subsection \ref{uina} the UV divergent parts of the parton distribution function (PDF) $f_i^{UnRenormalized}$ and the fragmentation function (FF) $D_i^{UnRenormalized}$ in the unrenormalized QCD appear as the $\frac{1}{\epsilon^n}$ poles in the limit $\epsilon \rightarrow 0$. Eq. (\ref{sgu}) guarantees that the $\frac{1}{\epsilon^n}$ poles in the limit $\epsilon \rightarrow 0$ in the partonic cross section ${\hat \sigma}^{UnRenormalized}$ in the unrenormalized QCD exactly cancel with the $\frac{1}{\epsilon^n}$ poles in the limit $\epsilon \rightarrow 0$ in the $f_1^{UnRenormalized} \otimes f_2^{UnRenormalized} \otimes D_1^{UnRenormalized} \otimes ...\otimes D_n^{UnRenormalized}$ in unrenormalized QCD for the scattering process in eq. (\ref{h37mg}). This is similar to cancelation of infrared (IR) and collinear divergences where the uncanceled infrared (IR) and collinear divergences in the partonic cross section ${\hat \sigma}^{UnRenormalized}$ in the unrenormalized QCD exactly cancel with the infrared (IR) and collinear divergences in the gauge links in the $f_1^{UnRenormalized} \otimes f_2^{UnRenormalized} \otimes D_1^{UnRenormalized} \otimes ...\otimes D_n^{UnRenormalized}$ for the scattering process in eq. (\ref{h37mg}) \cite{n1,n2,n3}.

\section{ Advantages of Unrenormalized QCD over Renormalized QCD to Study Hadron Production at High Energy Colliders }\label{adv}

In this section we will discuss the advantages of the unrenormalized QCD over the renormalized QCD to study hadron production at high energy colliders.

\subsection{ Many Values of the QCD Running Coupling in Renormalized QCD }\label{rc}

From eq. (\ref{bc}) one finds that the renormalized running coupling $g_R(\mu)$ is a function of the unphysical mass scale $\mu$ which implies that there can be infinite number of values of the running coupling $g_R(\mu)$ depending on infinite number of values of the unphysical mass scale $\mu$. Since the QCD coupling $g_R(\mu)$ is not a physical observable in renormalized QCD, it is not desirable to extract many values of the $g_R(\mu)$ for many values of $\mu$ from the experimental measurement of the hadronic cross section $\sigma$. These non-fixed values of the renormalized running coupling $g_R(\mu)$ unnecessarily complicate the calculation of $f_i^{Renormalized},~ {\hat \sigma}^{Renormalized},~ D_i^{Renormalized}$ in renormalized QCD by bringing additional renormalization group equations, see subsection \ref{rgn}.

\subsection{ Single Value of the QCD Coupling Constant in Unrenormalized QCD }

The situation is better in unrenormalized QCD because the bare coupling constant $g$ does not depend on the unphysical mass scale $\mu$. Hence one finds that there is only a single value of the coupling constant $g$ in unrenormalized QCD. This single value of the coupling constant $g$ in unrenormalized QCD is preferred than the infinite number of values of the QCD running coupling $g_R(\mu)$ in renormalized QCD. For example, we will need to extract only a single value of the coupling constant $g$ from the experimental measurement of the hadronic cross section $\sigma$ in eq. (\ref{sgu}) in unrenormalized QCD instead of extracting infinite number of values of running coupling $g_R(\mu)$ depending on infinite number of values of the unphysical mass scale $\mu$ in renormalized QCD from eq. (\ref{sgu}).

Note that the single value of the coupling constant $g$ in unrenormalized QCD is not similar to the constant electric charge $e$ in classical Maxwell theory. This is because while $e$ is the electric charge of the electron in classical Maxwell theory but $g$ is not the color charge $q^a(t)$ of the quark in classical Yang-Mills theory \cite{yang}. The constant coupling $g$ is different from the time dependent color charge $q^a(t)$ of the quark in classical Yang-Mills theory. The relation between the constant coupling $g$ and the time dependent color charge $q^a(t)$ of the quark in classical Yang-Mills theory is given by \cite{n5}
\bea
g^2 = q_1^2(t) + q_2^2(t)+q_3^2(t)+q_4^2(t)+q^2_5(t)+q_6^2(t)+q_7^2(t)+q_8^2(t)
\label{gqc}
\eea
where $a=1,2,...,8$ are the color indices. Hence one should not expect that the constant coupling $g$ means Coulomb-like force in Yang-Mills theory. This is different from Maxwell theory where the constant electric charge $e$ generates Coulomb potential. The color potential (Yang-Mills potential) $A_\mu^a(x)$ produced from the color charge $q^a(t)$ of the quark at rest is given by \cite{n4}
\bea
\Phi^a(t,r)=A_0^a(t,r)=\frac{q^b(t-\frac{r}{c})}{r}[\frac{e^{g\int dr \frac{Q(t-\frac{r}{c})}{r}}-1}{g\int dr \frac{Q(t-\frac{r}{c})}{r}}]_{ab}
\label{pqc}
\eea
where $dr$ integration is an indefinite integration and
\bea
Q^{ab}(t)=f^{abc}q^c(t).
\label{qcb}
\eea
From eq. (\ref{pqc}) one finds that the color potential is not like Coulomb potential even if the coupling $g$ is a fixed constant.

\subsection{ Many Values of the Running Quark Mass in Renormalized QCD }

Note that our notation of quark mass $m$ is for a single flavor. There are 6 different flavors of quarks with different masses in Yang-Mills theory. As mentioned earlier, since the renormalized running quark mass $m_R(\mu)$ depends on the unphysical mass scale $\mu$ one finds that $m_R(\mu)$ is not a physical observable in renormalized QCD. In addition to this, since the renormalized running quark mass $m_R(\mu)$ is a function of the unphysical mass scale $\mu$ one finds that there are infinite number of values of the running quark mass $m_R(\mu)$ depending on the infinite number of values of the unphysical mass scale $\mu$ in renormalized QCD. Hence it is not desirable to extract infinite number of values of $m_R(\mu)$ depending on the infinite values of the unphysical mass scale $\mu$ from the experimental measurement of the hadronic cross section $\sigma$. These non-fixed values of the renormalized running quark mass $m_R(\mu)$ unnecessarily complicate the calculation of the $f_i^{Renormalized},~ {\hat \sigma}^{Renormalized}$ and $D_i^{Renormalized}$ in the renormalized QCD by bringing additional renormalization group equations, see subsection \ref{rgn}.

\subsection{ Single Value of the Quark Mass in Unrenormalized QCD }\label{cqm}

On the other hand the situation is better in unrenormalized QCD because in the unrenormalized QCD the bare quark mass $m$ is independent of the unphysical mass scale $\mu$. Hence the bare quark mass $m$ in unrenormalized QCD is preferred than the infinite number of values of the QCD running quark mass $m_R(\mu)$ in renormalized QCD. This is because we will need to extract only a single value of the bare quark mass $m$ from the experimental measurement of the hadronic cross section $\sigma$ in eq. (\ref{sgu}) in unrenormalized QCD instead of extracting infinite number of values of running quark mass $m_R(\mu)$ depending on infinite number of values of the unphysical mass scale $\mu$ in renormalized QCD from eq. (\ref{sgu}).

\subsection{ Renormalization Group Equations are Additional Constraints in Renormalized QCD}\label{rgn}

As discussed in subsection \ref{rge} the hadron production cross section $\sigma$ in eq. (\ref{sgu}) at high energy colliders does not depend on the unphysical mass scale $\mu$ in renormalized QCD because the hadron is a physical observable in QCD \cite{css}. However, the partonic cross section ${\hat \sigma}^{Renormalized}$, the parton distribution function $f_i^{Renormalized}$ and the fragmentation function $D_i^{Renormalized}$ which are not physical observable in QCD depend on the unphysical mass scale $\mu$ in renormalized QCD. Hence in order to make sure that the physical observable (the hadronic cross section) $\sigma$ in eq. (\ref{sgu}) in renormalized QCD is independent of the unphysical mass scale $\mu$ one finds that the additional constraint equations (the renormalization group equations) are necessary. The renormalization group equations are given by
\bea
\mu \frac{d \alpha_s(\mu)}{d\mu} =\beta(\alpha_s)=-\beta_0 \alpha_s^2(\mu) -\beta_1 \alpha_s^3(\mu)-\beta_2 \alpha_s^4(\mu)-...
\label{rge3}
\eea
and
\bea
\mu \frac{d m_R(\mu)}{d\mu} =-m_R(\mu) \gamma(\alpha_s)=-\gamma_1 m_R(\mu)\alpha_s(\mu) -\gamma_2 m_R(\mu)\alpha_s^2(\mu)-\gamma_3 m_R(\mu)\alpha_s^3(\mu)-...\nonumber \\
\label{rge4}
\eea
for the running QCD coupling $\alpha_s(\mu)$ and the running quark mass $m_R(\mu)$ where $\beta$ and $\gamma$ are the beta function and the anomalous dimension respectively.

\subsection{ Renormalization Group Equations Are Not Necessary in Unrenormalized QCD }\label{rgun}

In unrenormalized QCD we have the bare coupling constant $g$ and the bare quark mass $m$ which are independent of the unphysical mass scale $\mu$. Hence the unphysical mass scale $\mu$ is absent in unrenormalized QCD. This is a major advantage of unrenormalized QCD over the renormalized QCD because in unrenormalized QCD the partonic cross section ${\hat \sigma}^{UnRenormalized}$, the parton distribution function $f_i^{UnRenormalized}$ and the fragmentation function $D_i^{UnRenormalized}$ are independent of the unphysical mass scale $\mu$ which automatically guarantee that the hadronic cross section $\sigma$ in unrenormalized QCD in eq. (\ref{sgu}) is independent of the unphysical mass scale $\mu$. Hence, unlike renormalized QCD where the additional constraint equations (the renormalization group equations, see eqs. (\ref{rge3}) and (\ref{rge4})) are necessary, no such renormalization group equations are necessary in unrenormalized QCD to study the same finite hadron production cross section $\sigma$ in eq. (\ref{sgu}) at high energy colliders. This is an enormous simplification in unrenormalized QCD over the renormalized QCD.

\subsection{ $\Lambda_{\rm QCD}$ is An Additional Unknown Parameter in Renormalized pQCD }

Note that when the renormalization group equation (\ref{rge3}) is solved one needs to know the integration constant. This integration constant is known as the $\Lambda_{QCD}$ which is an unknown parameter in renormalzied pQCD. This parameter is not present in the original Yang-Mills lagrangian \cite{yang}. Also this parameter $\Lambda_{QCD}$ can not be calculated by using renormalized pQCD because $\Lambda_{QCD}$ depends on coupling in fixed order pQCD calculation which in turn depends on $\Lambda_{QCD}$.

\subsection{ $\Lambda_{\rm QCD}$ is Absent in Unrenormalized QCD }

The unrenormalized QCD is much better than renormalzied QCD in this respect because there is no $\Lambda_{QCD}$ in unrenormalized QCD. This is because
in unrenormalized QCD the bare coupling constant $g$ is independent of the unphysical mass scale $\mu$. This means there is no renormalization group equation and hence there is no integration constant (the $\Lambda_{QCD}$) in the unrenormalized QCD. Hence, unlike the renormalized QCD where one has to extract this unknown parameter $\Lambda_{QCD}$ from the experimental data of the hadronic cross section $\sigma$ at high energy colliders, there is no such unknown parameter in unrenormalized QCD. This is a major advantage in the unrenormalized QCD over the renormalized QCD.

\subsection{ Renormalization Scheme Dependence in Renormalized pQCD }

Another problem which arises in renormalized fixed order pQCD calculation is the renormalization scheme dependence of the physical observable. This should not be the case because the physical observable should be renormalization scheme independent. However, explicit calculations at higher orders in coupling $g_R(\mu)$ in renormalized fixed order pQCD results have found the renormalization scheme dependence of the physical observable. Hence this remains one of the problem in fixed order renormalized pQCD calculation.

\subsection{ Renormalization Scheme Independence in Unrenormalized QCD }

In unrenormalized QCD there is no problem of renormalization scheme dependence because there is no renormalization. Since there is no renormalization scheme dependence in unrenormalized QCD, the unrenormalized QCD has advantage over the renormalized QCD.

\subsection{ Finite QCD Coupling Constant $g$ in Unrenormalized QCD Is The Same $g$ That Appears in Classical Yang-Mills Theory }

The lagrangian density ${\cal L}(x)$ in classical Yang-Mills theory is given by
\bea
&& {\cal L}(x) = {\bar \psi}(x)[i\gamma^\lambda {\overrightarrow \partial}_\lambda -m +g\gamma^\lambda T^d A_\lambda^d(x) ]\psi(x)
-\frac{1}{4}F_{\mu \nu}^d(x)F^{\mu \nu d}(x),\nonumber \\
&&F_{\mu \nu}^d(x)=\partial_\mu A_\nu^d(x)-\partial_\nu A_\mu^d(x)+gf^{dcb}A_\mu^c(x) A_\nu^b(x)
\label{lym}
\eea
where $g$ is the coupling constant and $m$ is the mass of the quark in classical Yang-Mills theory. As mentioned above the QCD coupling constant in unrenormalized QCD is finite and unique. Note that the QCD lagrangian density is obtained from the Yang-Mills lagrangian density from eq. (\ref{lym}). Hence one finds that the finite and unique QCD coupling constant $g$ in the unrenormalized QCD is the same finite coupling constant $g$ that appears in the classical Yang-Mills theory. Since the unrenormalized QCD coupling constant $g$ is finite, the unrenormalized partonic level cross section ${\hat \sigma}^{UnRenormalized}$, the unrenormalized parton distribution function (PDF) $f_i^{UnRenormalized}$ and the unrenormalized fragmentation function (FF)  $D_i^{UnRenormalized}$ become ultra violet (UV) divergent. The eq. (\ref{sgu}) guarantees that the ultra violet (UV) divergence in the unrenormalized partonic level cross section ${\hat \sigma}^{UnRenormalized}$ exactly cancels with the ultra violet (UV) divergence in the $f_1^{UnRenormalized} \otimes f_2^{UnRenormalized} \otimes D_1^{UnRenormalized} \otimes ...\otimes D_n^{UnRenormalized}$ making the hadronic cross section $\sigma$ finite.

\subsection{ Finite Quark Mass $m$ in Unrenormalized QCD Is The Same $m$ That Appears in Classical Yang-Mills Theory }

The quark mass $m$ appears in the classical Yang-Mills lagrangian density in eq. (\ref{lym}).
As mentioned above the quark mass in unrenormalized QCD is finite and unique. Note that the QCD lagrangian density is obtained from the Yang-Mills lagrangian density from eq. (\ref{lym}). Hence one finds that the finite and unique quark mass $m$ in the unrenormalized QCD is the same finite quark mass $m$ that appears in the classical Yang-Mills theory. Since the unrenormalized quark mass $m$ is finite, the unrenormalized partonic level cross section ${\hat \sigma}^{UnRenormalized}$, the unrenormalized parton distribution function (PDF) $f_i^{UnRenormalized}$ and the unrenormalized fragmentation function (FF)  $D_i^{UnRenormalized}$ become ultra violet (UV) divergent. The eq. (\ref{sgu}) guarantees that the ultra violet (UV) divergence in the unrenormalized partonic level cross section ${\hat \sigma}^{UnRenormalized}$ exactly cancels with the ultra violet (UV) divergence in the $f_1^{UnRenormalized} \otimes f_2^{UnRenormalized} \otimes D_1^{UnRenormalized} \otimes ...\otimes D_n^{UnRenormalized}$ making the hadronic cross section $\sigma$ finite.

\subsection{ Finite Hadronic Cross Section At High Energy Colliders Is Regularization Choice (Or UV Regulator) Independent }

As mentioned above the QCD coupling constant $g$ is finite in unrenormalized QCD and the quark mass $m$ is finite in unrenormalized QCD. Hence the unrenormalized partonic level cross section ${\hat \sigma}^{UnRenormalized}$, the unrenormalized parton distribution function (PDF) $f_i^{UnRenormalized}$ and the unrenormalized fragmentation function (FF)  $D_i^{UnRenormalized}$ become UV divergent and in diagrammatic calculation become regularization choice (or UV regulator) dependent. However, the eq. (\ref{sgu}) guarantees that the ultra violet (UV) divergence in the unrenormalized partonic level cross section ${\hat \sigma}^{UnRenormalized}$ exactly cancels with the ultra violet (UV) divergence in the $f_1^{UnRenormalized} \otimes f_2^{UnRenormalized} \otimes D_1^{UnRenormalized} \otimes ...\otimes D_n^{UnRenormalized}$ making the hadronic cross section $\sigma$ finite which proves that the finite hadronic cross section $\sigma$ in eq. (\ref{sgu}) is regularization choice (or UV regulator) independent.

Hence one finds that, unlike QED where there is no confinement, the hadron production cross section due to confinement in QCD at high energy colliders is regularization choice (or UV regulator) independent.

\section{ One Can Do Renormalization in QCD To Study Hadron Production At High Energy Colliders But It Is Not Necessary }

As mentioned above, in the dimensional regularization in $4-2\epsilon$ dimensions, the quantities which diverge as $\epsilon \rightarrow 0$ in QCD are not physical observable. What matters is that the results of the physical observable are finite. The partonic cross section is not physical observable but hadronic cross section is physical observable. Since partonic cross section is not physical observable there is nothing wrong if the partonic cross section becomes infinite.

The workers who do perturbative calculations try to make partonic cross section finite by performing renormalization. However, as mentioned above since the partonic cross section is not physical observable it is not necessary to make the partonic cross section finite and hence it is not necessary to do renormalization to make the partonic cross section finite. The workers who do perturbative calculation can keep the ultra violet (UV) infinite partonic cross section by not doing renormalization. These ultra violet (UV) divergences in the partonic cross section will exactly cancel with all the ultra violet (UV) divergences in the parton distribution functions (PDFs) and in the fragmentation functions (FFs), predicting the finite hadronic cross section at high energy colliders which is proved in eq. (\ref{sgu}).

The main issue is to separate the ultra violet (UV) divergent part of the partonic cross section from the finite part of the partonic cross section. One can use any regularization scheme for this purpose. For example, if one uses the dimensional regularization in $4-2\epsilon$ dimensions then the ultra violet (UV) divergent part of the partonic cross section appears as the $\frac{1}{\epsilon^n}$ poles in the limit $\epsilon \rightarrow 0$. Similarly the ultra violet (UV) divergent parts of the parton distribution function (PDF) and the fragmentation function (FF) appear as the $\frac{1}{\epsilon^n}$ poles in the limit $\epsilon \rightarrow 0$. Eq. (\ref{sgu}) guarantees that the $\frac{1}{\epsilon^n}$ poles in the limit $\epsilon \rightarrow 0$ in the partonic cross section ${\hat \sigma}^{UnRenormalized}$ in the unrenormalized QCD exactly cancel with the $\frac{1}{\epsilon^n}$ poles in the limit $\epsilon \rightarrow 0$ in the $f_1^{UnRenormalized} \otimes f_2^{UnRenormalized} \otimes D_1^{UnRenormalized} \otimes ...\otimes D_n^{UnRenormalized}$ in unrenormalized QCD producing the finite hadronic cross section $\sigma$. This type of cancelation of the ultra violet (UV) divergences between ${\hat \sigma}^{UnRenormalized}$ and $f_1^{UnRenormalized} \otimes f_2^{UnRenormalized} \otimes D_1^{UnRenormalized} \otimes ...\otimes D_n^{UnRenormalized}$ in unrenormalized QCD is similar to the cancelation of infrared (IR) and collinear divergences between ${\hat \sigma}^{UnRenormalized}$
and $f_1^{UnRenormalized} \otimes f_2^{UnRenormalized} \otimes D_1^{UnRenormalized} \otimes ...\otimes D_n^{UnRenormalized}$ \cite{n1,n2,n3}.

Hence although workers who do perturbative QCD perform renormalization to calculate finite {\bf partonic cross section} but it is not necessary to do renormalization to calculate the finite {\bf hadronic cross section} at high energy colliders. The ultra violet (UV) divergences in $f_1^{UnRenormalized} \otimes f_2^{UnRenormalized} \otimes D_1^{UnRenormalized} \otimes ...\otimes D_n^{UnRenormalized}$ act as natural counter terms to the ultra violet (UV) divergences in ${\hat \sigma}^{UnRenormalized}$ in the unrenormalized QCD.

This implies that, as far as renormalization is concerned, the QCD seems to be a better theory than QED.

Note that, we are not saying that one should not do renormalization in pQCD to study hadron production at high energy colliders. What we are saying is that although one can do renormalization in pQCD to study hadron production at high energy colliders but it is not necessary to renormalize QCD to study hadrons production at high energy colliders. One can get exactly the same cross section for the hadron production at high energy colliders by using unrenormalized QCD which we have proved in eq. (\ref{sgu}).

\section{Conclusions}
In this paper by using the path integral formulation of the background field method of QCD in the presence of SU(3) pure gauge background field we have simultaneously proved the renormalization of ultra violet (UV) divergences and the factorization of infrared (IR) and collinear divergences in QCD at all orders in coupling constant. We have proved that although the perturbative QCD is renormalizable but due to confinement in QCD it is not necessary to renormalize QCD to study hadrons production at high energy colliders. This is consistent with the fact that the partons are not directly experimentally observed but the hadrons are directly experimentally observed.


\begin{thebibliography}{99}

\bibitem{hf} G. 't Hooft and M.J.G. Veltman, Nucl.Phys. B44 (1972) 189.

\bibitem{gs} D. J. Gross and F. Wilczek, Phys. Rev. Lett. 30 (1973) 1343; D. Politzer, Phys. Rev. Lett. 30 (1973) 1346.

\bibitem{s1} J. C. Collins, D. E. Soper and G. Sterman, Nucl. Phys. B223 (1983) 381; 261 (1985)
104.

\bibitem{s2} J. Collins, D. E. Soper and G. Sterman, Phys. Lett. 109B (1982) 388; 126B (1983) 275; 134B (1984) 263.

\bibitem{s3} G. C. Nayak, J-W. Qiu and G. Sterman, Phys. Lett. B613 (2005) 45; Phys.Rev. D72 (2005) 114012; Phys.Rev. D74 (2006) 074007.

\bibitem{n1} G. C. Nayak, Eur. Phys. J. C76 (2016) 448.

\bibitem{n2} G. C. Nayak, Phys. Part. Nucl. Lett. 13 (2016) 417.

\bibitem{n3} G. C. Nayak, Phys. Part. Nucl. Lett. 14 (2017) 18.

\bibitem{gram} G. Grammer and D. R. Yennie, Phys. Rev. D8 (1973) 4332.

\bibitem{n4} G. C. Nayak, JHEP 1303 (2013) 001.

\bibitem{n5} G. C. Nayak, Eur. Phys. J. C73 (2013) 2442.

\bibitem{ab} L. F. Abbott, Nucl. Phys. B185 (1981) 189.

\bibitem{nlsz} G. C. Nayak, arXiv:1705.02927 [hep-ph].

\bibitem{alp} S. Bethke, Eur. Phys. J. C 64 (2009) 689.

\bibitem{yang} C. N. Yang and R. Mills, Phys. Rev. 96 (1954) 191.

\bibitem{css} J. C. Collins, D. E. Soper and G. Sterman, Adv. Ser. Direct High Energy Phys. 5 (1988) 1, hep-ph/0409313.


\end{thebibliography}
\end{document}